\documentclass[11pt,twoside]{article}
\usepackage{exscale}
\usepackage{amssymb}
\usepackage[dvips]{graphicx}
\usepackage{epsfig}
\usepackage{latexsym}
\usepackage{floatflt}

\author{Pawe{\l} Cieciel\c{a}g, Micha{\l} Chodorowski and Andrzej Kudlicki}
\title{\textbf{Local Group Velocity Versus Gravity: \\
		Nonlinear Effects}}

\begin{document}

\newcommand{\om}{$\Omega_{0}$}
\newcommand{\hub}{$H_{0}$}
\newcommand{\vg}{$\mathbf{v} \cdot \mathbf{g}$}
\newcommand{\geom}{\mathbf{\frac{x'-x}{|x'-x|^3}}}
\newcommand{\dotdeg}{\hbox{$.\!\!^\circ$}}
\newcommand{\vecv}{\mathbf{v}}
\newcommand{\vecg}{\mathbf{g}}
\newcommand{\vecx}{\mathbf{x}}
\newcommand{\kms}{$\hbox{km}\cdot \hbox{s}^{-1}$}
\newcommand{\hmpc}{$h^{-1}\,\hbox{Mpc}$}
\newcommand{\mhmpc}{{\, h^{-1}\rm Mpc}}

\newcommand{\done}{\delta^{(1)}}
\newcommand{\de}{\delta}
\newcommand{\te}{\theta}
\newcommand{\la}{\lambda}
\newcommand{\p}{\partial}
\newcommand{\f}{\frac}
\newcommand{\ap}{\approx}
\newcommand{\Om}{\Omega}
\newcommand{\w}{\omega}
\newcommand{\s}{\sigma}
\newcommand{\al}{\alpha}
\newcommand{\fd}{\tilde{\delta}}
\newcommand{\fv}{\tilde{v}}
\newcommand{\fJ}{\tilde{J}}
\newcommand{\fW}{\widetilde{W}}
\newcommand{\bfx}{{\bf x}}
\newcommand{\bfr}{{\bf r}}
\newcommand{\bfs}{{\bf s}}
\newcommand{\bft}{{\bf t}}
\newcommand{\bfz}{{\bf z}}
\newcommand{\bfy}{{\bf y}}
\newcommand{\bfk}{{\bf k}}
\newcommand{\bfv}{{\bf v}}
\newcommand{\bfq}{{\bf q}}
\newcommand{\bfg}{{\bf g}}
\newcommand{\scg}{\tilde{\bf g}}
\newcommand{\bfp}{{\bf p}}
\newcommand{\bfu}{{\bf u}}
\newcommand{\vr}{{\varrho}}
\newcommand{\calF}{{\cal F}}
\newcommand{\calO}{{\cal O}}
\newcommand{\calQ}{{\cal Q}}
\newcommand{\calC}{{\cal C}}
\newcommand{\calI}{{\cal I}}
\newcommand{\calL}{{\cal L}}
\newcommand{\calK}{{\cal K}}
\newcommand{\calN}{{\cal N}}
\newcommand{\calS}{{\cal S}}
\newcommand{\calH}{{\cal H}}
\newcommand{\calP}{{\cal P}}
\newcommand{\eps}{{\epsilon}}
\newcommand{\bc}{\begin{center}}
\newcommand{\be}{\begin{equation}}
\newcommand{\ee}{\end{equation}}
\newcommand{\ec}{\end{center}}
\newcommand{\lan}{\langle}
\newcommand{\ran}{\rangle}
\newcommand{\sig}{\sigma_{_{\!{\!G}}}}
\newcommand{\mr}{\overline\vr}
\newcommand{\lra}{{\leftrightarrow}}
\newcommand{\hP}{{\hat P}}
\newcommand{\hl}{\vspace{0.2cm}\hrule width \hsize height 0.45pt\vspace{0.2cm} }

\newcommand{\spose}[1]{\hbox to 0pt{#1\hss}}
\newcommand{\lta}{\mathrel{\spose{\lower 3pt\hbox{$\mathchar"218$}}
 \raise 2.0pt\hbox{$\mathchar"13C$}}}
\newcommand{\gta}{\mathrel{\spose{\lower 3pt\hbox{$\mathchar"218$}}
 \raise 2.0pt\hbox{$\mathchar"13E$}}}

\newcommand{\dd}{\,{\rm d}}
\newcommand{\ie}{{\it i.e.},\,}
\newcommand{\etal}{{\it et al.\ }}
\newcommand{\eg}{{\it e.g.},\,}
\newcommand{\cf}{{\it cf.\ }}
\newcommand{\vs}{{\it vs.\ }}
\newcommand{\zdot}{\makebox[0pt][l]{.}}
\newcommand{\up}[1]{\ifmmode^{\rm #1}\else$^{\rm #1}$\fi}
\newcommand{\dn}[1]{\ifmmode_{\rm #1}\else$_{\rm #1}$\fi}
\newcommand{\upd}{\up{d}}
\newcommand{\uph}{\up{h}}
\newcommand{\upm}{\up{m}}
\newcommand{\ups}{\up{s}}
\newcommand{\arcd}{\ifmmode^{\circ}\else$^{\circ}$\fi}
\newcommand{\arcm}{\ifmmode{'}\else$'$\fi}
\newcommand{\arcs}{\ifmmode{''}\else$''$\fi}
\newcommand{\MS}{{\rm M}\ifmmode_{\odot}\else$_{\odot}$\fi}
\newcommand{\RS}{{\rm R}\ifmmode_{\odot}\else$_{\odot}$\fi}
\newcommand{\LS}{{\rm L}\ifmmode_{\odot}\else$_{\odot}$\fi}

\newcommand{\Abstract}[1]{{\footnotesize\begin{center}ABSTRACT\end{center}
\vspace{1mm}\par#1\par}}

\newenvironment{references}%
{
\frenchspacing
\renewcommand{\thesection}{}
\renewcommand{\in}{{\rm in }}
\renewcommand{\AA}{Astron.\ Astrophys.}
\newcommand{\AAS}{Astron.~Astrophys.~Suppl.~Ser.}
\newcommand{\ApJ}{Astrophys.\ J.}
\newcommand{\ApJS}{Astrophys.\ J.~Suppl.~Ser.}
\newcommand{\ApJL}{Astrophys.\ J.~Letters}
\newcommand{\AJ}{Astron.\ J.}
\newcommand{\IBVS}{IBVS}
\newcommand{\PASP}{P.A.S.P.}
\newcommand{\Acta}{Acta Astron.}
\newcommand{\MNRAS}{MNRAS}
\newcommand{\and}{{\rm and }}
\section{{\rm REFERENCES}}
\sloppy \hyphenpenalty10000
\begin{list}{}{\leftmargin1cm\listparindent-1cm
\itemindent\listparindent\parsep0pt\itemsep0pt}}%
{\end{list}\vspace{2mm}}

\def\TYLDA{~}
\newlength{\DW}
\settowidth{\DW}{0}
\newcommand{\dw}{\hspace{\DW}}

\newcommand{\refitem}[5]{\item[]{#1} #2%
\def\REFARG{#3}\ifx\REFARG\TYLDA\else, {\it#3}\fi
\def\REFARG{#4}\ifx\REFARG\TYLDA\else, {\bf#4}\fi
\def\REFARG{#5}\ifx\REFARG\TYLDA\else, {#5}\fi.}

\newcommand{\Section}[1]{\section{#1}}
\newcommand{\Subsection}[1]{\subsection{#1}}
\newcommand{\Acknow}[1]{\par\vspace{5mm}{\bf Acknowledgements.} #1}
\pagestyle{myheadings}

\newfont{\bb}{ptmbi at 12pt}
\newcommand{\xrule}{\rule{0pt}{2.5ex}}
\newcommand{\xxrule}{\rule[-1.8ex]{0pt}{4.5ex}}
\def\thefootnote{\fnsymbol{footnote}}

\maketitle

\Abstract{
We use numerical simulations to study the relation between the
velocity of the Local Group (LG) and its gravitational acceleration.
This relation serves as a test for the kinematic origin of the CMB
dipole and as a method for estimating $\beta$. We calculate the
misalignment angle between the two vectors and compare it to the
observed value for the PSCz survey. The latter value is near the upper
limit of the $95$ \% confidence interval for the angle; therefore, the
nonlinear effects are unlikely to be responsible for the whole
observed misalignment. We also study the relation between the
amplitudes of the LG velocity and gravity vectors. In an $\Omega = 1$
Universe, the {\em smoothed} gravity of the LG turns out to be a
biased low estimator of the LG (unsmoothed) velocity. In an $\Omega =
0.3$ Universe, the estimator is biased high. The discussed biases are,
however, only a few per cent, thus the linear theory works to good
accuracy. The gravity-based estimator of the LG velocity has also a
scatter that limits the precision of the estimate of $\beta$ in the
LG velocity--gravity comparisons. The random error of $\beta$ due to
nonlinear effects amounts to several per cent.
}

\Section{Introduction}
\label{sec:intro}
The dipole anisotropy of the Cosmic Microwave Background (CMB)
temperature is widely believed to reflect, via the Doppler shift, the
motion of the Local Group (LG) with respect to the CMB rest
frame. When transformed to the barycenter of the LG, this motion is
towards $(l,b) = (276^{\circ} \pm 3^{\circ},30^{\circ} \pm
2^{\circ})$, and of amplitude $v_{\rm LG} = 627 \pm 22$ \kms, as
inferred from the 4-year COBE data (Lineweaver \etal 1996).
Alternative models which assume that the dipole is due to a metric
fluctuation (e.g., Paczy\'nski \& Piran 1990) have problems with
explaining its observed achromaticity and the relative smallness of
the CMB quadrupole.

An additional argument in favour of the kinematic interpretation of
the CMB dipole is its remarkable alignment with the LG acceleration,
inferred from galaxy distribution. The acceleration on the LG (i.e.,
the galaxy dipole), inferred from the {\em IRAS\/} 1.2 Jy survey
points $\sim 25^\circ$ away from the CMB dipole (Strauss \etal 1992;
hereafter S92). The recently completed {\em IRAS\/} PSCz survey
allowed to make further progress on this topic. Schmoldt \etal (1999;
hereafter S99) found the PSCz galaxy dipole to be within $15^\circ$ of
the CMB dipole. Rowan-Robinson \etal (1999), performing a similar
analysis of the galaxy dipole but out to a larger distance $300$
\hmpc, obtained the misalignment angle as small as $13^\circ$.

In the linear theory, 
(e.g., Peebles 1980)
the peculiar velocity of any galaxy, $\bfv$, is
directly proportional to its gravitational acceleration, $\bfg$: $\bfv
= (2f / 3H\Omega) \bfg\,$. Here, $\Omega$ denotes the cosmic
{\em matter} density parameter, $H$ is the Hubble
constant and

\be
f(\Omega,\Lambda) \simeq \Omega^{0.6} + \f{\Lambda}{70} \left(1 +
\f{\Omega}{2}\right) 
\label{eq:f_fac}
\ee 
(Lahav \etal 1991; note a very weak dependence on the cosmological
constant). 

The acceleration is caused by the gravitational pull of the
surrounding matter, \be \bfg(\bfr) = G \rho_b \int {\rm d}^3 r'
\de(\bfr') \f{\bfr'- \bfr}{\vert \bfr'- \bfr \vert^3} \,.
\label{eq:g}
\ee 
Here, $G$ is the gravitational constant, $\rho_b$ is the background
density, and $\de$ is the {\em mass\/} density fluctuation
field.
If we define the {\em scaled\/} gravity, $\scg 
\equiv (2f / 3H\Omega) \bfg$, then in the linear theory
\be
\bfv = \scg \,.
\label{eq:v-sg}
\ee
The vector $\scg$ can be measured from redshift surveys which give an
estimate of the three-dimensional {\em galaxy\/} density field,
$\de_g$. In the simplest biasing model the galaxy and mass
distributions are linearly related, $\de_g = b \de$, where $b$ is a
linear biasing factor. The scaled gravity is then

\be \scg(\bfr) = \beta \int \f{{\rm d}^3 r'}{4\pi}
\de_g(\bfr') \f{\bfr'- \bfr}{\vert \bfr'- \bfr \vert^3} \,,
\label{eq:g_scaled}
\ee 
where $\beta \equiv f/b$ and we express distances in units of \kms.

The amplitude of the scaled gravity depends on $\beta$. Therefore, a
comparison between the LG scaled gravity and the CMB dipole can
serve not only as a test for the kinematic origin of the latter but
also as a measure of the parameter $\beta$.  Combined with other
constraints on bias, it may yield an estimate of $\Omega$ itself.

However, the estimate~(\ref{eq:g_scaled}) of the LG velocity from a
particular redshift survey will in general differ from its true
velocity, for a number of reasons. We enumerate them here following
S99: 

\begin{itemize} 

\item The finite volume of the survey -- the galaxy dipole may be
affected by contributions from outside;

\item Unsurveyed regions within the volume;

\item The shot noise -- the galaxy density field is sampled
discretely, at galaxy's positions;

\item Redshift-space distortions -- in redshift surveys the radial
coordinate is not the true distance to a galaxy but rather a sum of
the distance and the radial component of its a priori unknown peculiar
velocity;

\item Nonlinear effects -- the scaled acceleration is equal to the
velocity only in the linear regime, while redshift surveys unveil
nonlinear structure of the galaxy distribution.

\end{itemize}
In the proper process of the LG gravity--velocity comparison, all
these effects should be accounted for. In the present paper, we will
concentrate on the nonlinear effects (hereafter NE). In general, the
NE can be due to both nonlinear gravity and nonlinear biasing. Here,
we will only consider the effects of nonlinear gravity, and we will
use the term `NE' in this, more narrow, meaning.

NE modify the linear relation~(\ref{eq:v-sg}) in a number of ways.
Non-local nature of gravity, somewhat hidden at linear order,
manifests itself at higher orders. The relation between the velocity
and the scaled gravitational acceleration becomes not only non-linear
but also non-local, so at a given point it has a scatter. The NE may
also spoil the alignment between the two vectors. Thus, the NE may
influence not only the accuracy of estimating $\beta$ from Local Group
dipole comparisons, but also its precision. In this paper we address
both issues.

A common method of constraining cosmological parameters by the LG
gravity--velocity comparison is to apply a maximum-likelihood
analysis. In a given cosmological model, one maximizes the likelihood
of measuring the scaled gravity of the LG given the true
(CMB-inferred) value of the LG velocity (Juszkiewicz, Vittorio \& Wyse
1990, Lahav, Kaiser \& Hoffman 1990, S92, S99). In such an analysis, a
proper object describing the nonlinear effects is the decoherence
function, i.e. the cross-correlation coefficient of the Fourier modes
of the gravity and velocity fields (S92). 

Here, we will study the NE in real space. In particular, we will study
the evolution of the misalignment angle, and the relation between the
amplitudes of the gravity and velocity vectors. This approach is more
appropriate in the case of the `numerical analysis' of S99, where they simply
equated the LG gravity, inferred from the {\it IRAS\/} PSCz catalog,
to the LG velocity. To account for the NE (and other effects mentioned
earlier), S99 used mock catalogs to compute the ratio between the
reconstructed gravity dipole at the observer's position and its true
N-body velocity. Then they used the average of the ratios calculated
from mock catalogs as a multiplicative factor which
relates the reconstructed gravity dipole to the real LG velocity. The
mock catalogs were constructed for different cosmologies. In our
opinion, if the factor varies systematically from model to model,
averaging it makes no sense. Rather, the results should be expressed
explicitly in a model-dependent way. (This is the case of the
likelihood analysis of S99.) One of the goals of the present paper is
to clarify this issue.

The NE in real space were studied by means of N-body simulations by
Davis, Strauss \& Yahil (1991); however, only in a standard CDM
model. Here we investigate the effects of varying $\Omega$ on the NE,
also using numerical simulations. Instead of using a N-body scheme, we
model cold dark matter as a pressureless cosmic fluid (see Peebles
1987). The outline of this paper is as follows: In
Section~\ref{sec:sim} we present our numerical model of the large
scale structure evolution and how we select Local Group candidates in
the data.  Next we discuss the differences between the Local Group
velocity and acceleration: in Section~\ref{sec:mis} we discuss the
angle between the two vectors, in Section~\ref{sec:am} -- their
amplitudes.  We summarize our results in Section~\ref{sec:sum}.

\Section{The simulations}

\label{sec:sim}
\Subsection{The Numerical Model}
\label{sec:code}
We have performed our simulations using an Eulerian, uniform-grid based 
code -- CPPA (Cosmological Pressureless Parabolic Advection, see
Kudlicki, Plewa \& R\'o\.zyczka 1996, Kudlicki \etal 2000a, 
Kudlicki \etal 2000c). The main features of CPPA are parabolic density 
and velocity profiles, variable timestep, periodic boundary conditions
and a flux interchange procedure, implemented as
an approximation to the solution of the Boltzmann equation.
\label{sec:mod}
We have studied two cosmological models with $\Omega=1$ and $\Omega=0.3$,
assuming Gaussian initial conditions. For better statistics we performed 4 
realizations of each of the models, varying random phases of 
the initial density field.
The grid we used was $64^3$, and the comoving size of the
simulation box was $100$ \hmpc\ on a side. 

The linear velocity depends on the cosmological constant
very weakly (see eq.~\ref{eq:f_fac}); this holds also for higher
orders (see Appendix B.3 of Scoccimarro \etal 1998 and Nusser \&
Colberg 1998), so we were free to assume $\Lambda = 0$
in all our models.
To make the simulated gravity of the LG as close as
possible to that inferred from the {\em IRAS\/} PSCz survey, for the
mass power spectrum we adopted the power spectrum of the PSCz galaxies
(Sutherland \etal 1999):
\begin{eqnarray}
\label{eq:power}
P(k) & = & \frac{Bk}{\lbrace 1+[ak+(bk)^{3/2}+(ck)^2]^{\nu})^{2/\nu}\rbrace} \\
a & = & 6.4/\Gamma \mhmpc\,,~~~b=3.0/\Gamma \mhmpc\,, \nonumber\\
c & = & 1.7/\Gamma \mhmpc\,,~~~\nu=1.13 \nonumber
\end{eqnarray}
with $\Gamma = 0.2$ as best fitted value.  To normalize the power
spectrum we used the observed local abundance of galaxy clusters. The
present value of $\s_8$, $\s_{8,0}$, is a function of $\Omega$ and for
the case of $\Lambda = 0$, considered here, it is given by the
relation (Eke, Cole \& Frenk 1996)
\be
\s_{8,0} = (0.52 \pm 0.04) \Omega^{-0.46 + 0.10 \Omega} \,.
\label{eq:sigma_8}
\ee 
This relation changes only slightly with the shape of the power
spectrum. It is also very similar for the case $\Lambda = 1 -
\Omega$.
For $\Omega = 1$, $\s_{8,0} \simeq 0.52$, while for $\Omega =
0.3$, $\s_{8,0} \simeq 0.87$. 

\Subsection{Selection of LG candidates}
\label{sec:lg}
To study the relation between the velocity and the gravity of the LG,
from the simulation grid we selected the cells which have properties
resembling those of the LG. We chose these `LG candidates' following
the criteria presented in G\'orski \etal (1989), and used by Davis,
Strauss \& Yahil (1991): the peculiar velocity of the candidate must
be $|\bfv| = 600 \pm 10$ \kms, the candidate must be in a region in
which the fractional overdensity, averaged in a radius $R = 500$ \kms,
$\de^R$, is in the range $-0.2<\de^R < 1.0$, and the cosmic flow
within this volume is quiet: $|\bfv - \bfv^R| < 0.3|\bfv|$, where
$\bfv^R$ is the mean velocity within the averaging sphere. In general,
the symbol $x^R$ denotes the quantity $x$ averaged with a top-hat
filter of radius $R$. The chosen value $600$ \kms\ is not exactly
equal to the present estimate of the LG velocity, given in
Section~\ref{sec:intro}.  However, for the purposes of the present
paper the difference between the two is not significant, while the
value $600$ \kms\ facilitates a visual analysis of the results.

\Section{The misalignment angle}

\label{sec:mis}
\Subsection{Analytic predictions}
\label{subsec:qp}
As stated in Section~\ref{sec:intro}, the NE spoil the alignment
between the velocity and gravity vectors. In this Section we study the
evolution of their misalignment angle, $\alpha$, given by
\be 
\cos \alpha = \frac{\bfv \cdot \bfg}{vg}
\,.
\label{eq:alpha}
\ee
In the linear regime, $\bfv$ is proportional to $\bfg$ and the misalignment 
angle is exactly zero.
The linear velocity depends on $\Omega$ (and only weakly on $\Lambda$)
solely via the multiplicative factor $f$ (eq.~\ref{eq:f_fac}).
This is also approximately true for the nonlinear velocity (Appendix
B.3 of Scoccimarro \etal 1998 and Nusser \& Colberg
1998). Moreover, the gravity field scales linearly with $\Omega$ (see
eq.~\ref{eq:g}). As a result, in equation~(\ref{eq:alpha}) the factors
$f$ and $\Omega$ cancel out and we expect the misalignment angle to be
practically independent of $\Omega$.

If the fields are smoothed on mildly nonlinear scales,
the perturbation theory can be 
applied to predict qualitatively the
evolution of $\alpha$. Note first that the angle between the velocity
and the acceleration equals to the angle between the velocity and the
{\em scaled\/} acceleration (eq.~\ref{eq:g_scaled}). Expanding
the velocity and the scaled gravity in perturbative series, $\bfv =
\bfv_1 + \bfv_2 + \ldots$, $\scg = \scg_1 + \scg_2 +
\ldots$, and calculating the expectation value of the angle to the
leading order we obtain

\begin{equation}
\langle\alpha^2\rangle = \left\langle\frac{(\bfv_2-\scg_2)^2 - 
[\hat{\bfv}_1 \cdot (\bfv_2-\scg_2)]^2}{v_1^2}\right\rangle
\,.
\label{eq:alpha_2}
\end{equation}

Here, $\hat{\bfv}_1 = \bfv_1/v_1$, and the symbol $\lan \cdots \ran$
denotes ensemble averaging. Perturbative solutions for $\bfv_2$ and
$\scg_2$ (e.g., Goroff \etal 1986) yield the following 
approximate scaling
relations: $v_2 \sim \tilde{g}_2 \sim v_1 \de_1$. Since to
the leading order $\de_1 = \de$ and typically $\de \sim \s$,  we have

\be
\langle \alpha^2 \rangle^{1/2} \propto \s 
\,,
\label{eq:alpha_scal}
\ee
where $\s \equiv \langle \delta^2 \rangle^{1/2}$ is the r.m.s.
fluctuation of the mass density field. As described earlier, we expect
the coefficient of proportionality in the above relation to be
insensitive to $\Omega$.  On the other hand, we expect it to depend on
the relative amount of power on small (nonlinear) scales.  In
equation~(\ref{eq:alpha_2}), $\alpha$ is expressed in terms of only
first and second-order perturbative contributions to the velocity and
gravity fields. Despite this, an analytic calculation of the
misalignment angle is impossible, because it involves averaging of a
ratio of two non-Gaussian variables. In short, the resulting series
cannot be truncated.

Thus far, our analysis concerned the evolution of the misalignment
angle of the smoothed velocity and gravity fields. In practice, the
reconstruction of the LG gravitational acceleration from the
redshift-space galaxy field does involve smoothing (to mitigate strong
nonlinear effects, to reduce shot-noise and distance uncertainty,
etc.). The LG velocity, however, is inferred directly from the CMB
dipole, so it is not a subject to any smoothing, except the one needed
to reflect the finite size of the LG. As that smoothing, S92 adopt a
top-hat of radius $100$ \kms\ (1 \hmpc). Thus, while the LG
acceleration is smoothed, the LG velocity is (almost) unsmoothed.
Though in the linear regime $\bfv^R = \bfg^R$, if $\bfv
\ne \bfv^R$, then $\bfv \ne \bfg^R$. In other words, if 
some part of the central velocity is due to matter distribution within
the smoothing radius (what is quite natural to expect) we cannot
expect unsmoothed velocity to equal smoothed acceleration even in the
linear regime (see also Berlind, Narayanan \& Weinberg 2000). 

\Subsection{Numerical results}
\label{subsec:num}
We have calculated the misalignment angle
from our simulations. First we smoothed both velocity and gravity
fields with a top-hat of radius $500$ \kms . 
This radius is equal to the minimum radius of smoothing used
for the {\it IRAS\/} gravitational field calculations. As a
characteristic angle we adopted the quantity $\lan \alpha^2
\ran^{1/2}$, which, for simplictity, we will denote $\alpha$.

\begin{figure}[h]
\centerline{\includegraphics[angle=0,scale=0.6]{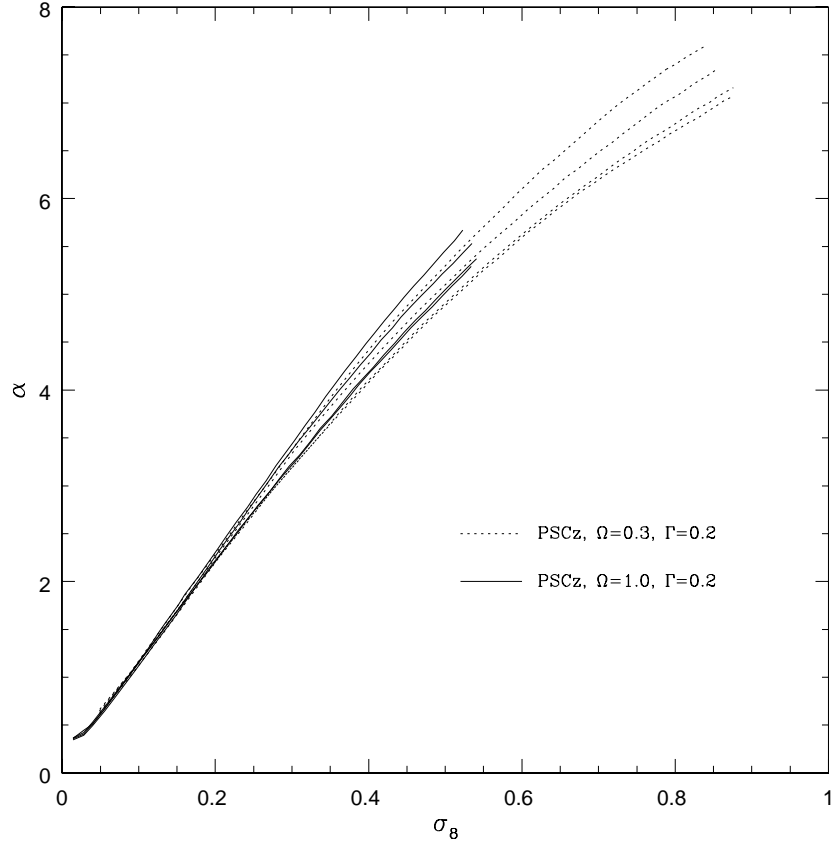}}
\caption{The characteristic angle between the smoothed velocity and
smoothed acceleration of all points, for models with $\Omega = 1$
(solid lines) and $\Omega = 0.3$ (dotted lines), as a function of the 
density dispersion in spheres of comoving radius $800$ \kms. 
\label{fig:angle}
}
\end{figure}

Figure~\ref{fig:angle} shows the \ time-evolution of $\alpha$ for all
points of the simulation as a function of the r.m.s. value of mass
fluctuations in spheres of canonical radius $800$ \kms, $\s_8$. The
angle depends linearly on $\s_8$ up to $\s_8 \simeq 0.5$. Moreover, it
is practically independent of $\Omega$: the systematic difference
between the values for models with the same phases of the initial
inhomogeneities and different $\Omega$ is not greater than the scatter
for models with the same $\Omega$ and different phases.

Our second step is to apply different smoothings to the {\em final\/}
gravity and velocity vectors, to mimic better real situation. 
The gravitational acceleration is smoothed with a filter described
earlier, but the velocity should now be smoothed with a top-hat of
radius $100$ \kms\ (to reflect the finite size of the LG).  
Since the cell volume in our simulations corresponds to that of a 
sphere with radius of 97 \kms, the unsmoothed velocity field is used 
for simplicity.

\begin{figure}[h]
\centerline{\includegraphics[angle=270,scale=0.6]{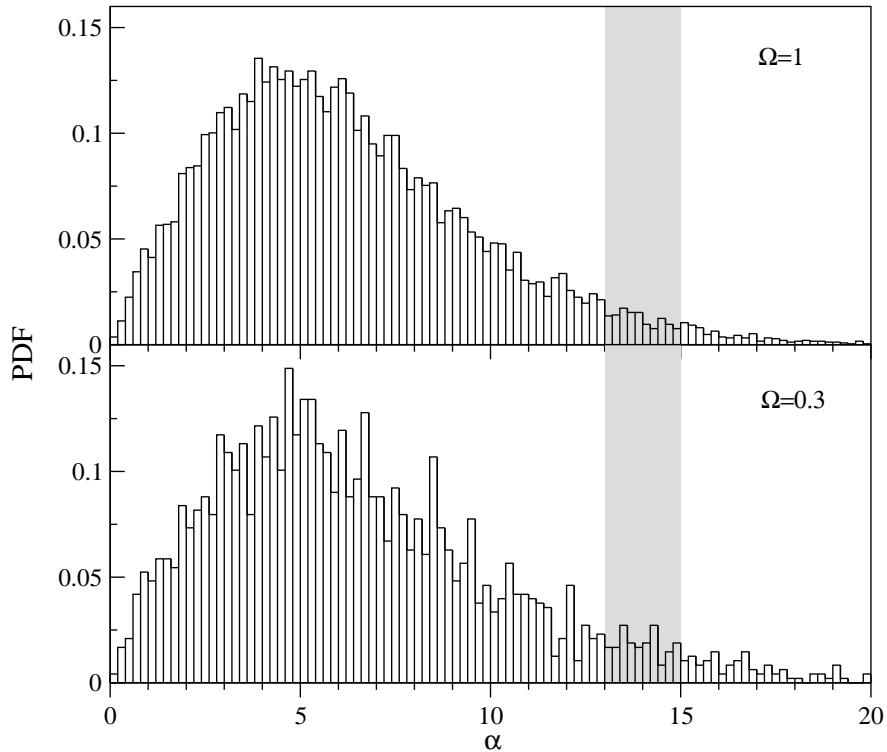}}
\caption{The simulated PDF of the misalignment angle between velocity and
smoothed gravity for LG candidates.
Upper panel: $\Omega=1.0$; lower: $\Omega=0.3$.
Shaded area represents observational constraints from the PSCz catalogue.
\label{fig:angle_LG}
}
\end{figure}

Figure~\ref{fig:angle_LG} shows distributions of
$\alpha$ for different $\Omega$ models with the above
normalizations. The mean values of the characteristic angle are
$7.0^\circ$ for $\Omega = 1$ and $7.7^\circ$ for $\Omega = 0.3$. On
the confidence level of 95 \%, $\alpha$ is smaller than respectively
$12.6^\circ$ and $13.9^\circ$ in the model with $\Omega=1$
and $\Omega=0.3$. Thus, the measured misalignment angle between the LG
velocity and PSCz galaxy dipole ($13$--$15^\circ$) is rather unlikely
to be caused entirely by nonlinear effects. Other effects must play a
role here as well, like, e.g., incomplete sky coverage and shot noise.

The values of the angle we obtained are smaller than the value
obtained by Davis \etal (1991) from N-body simulations,
$10.2^\circ$. However, we use the spectrum of the PSCz galaxies,
while the power spectrum they used was a standard CDM. This model has
more power on small (nonlinear) scales, so we expect a larger effect in
this case. We additionally calculated the angle in a standard CDM
model (the power spectrum given by eq.~\ref{eq:power} with $\Gamma = 0.5$). 
The mean angle indeed increased from $7.0^\circ$ to $8.1^\circ$.
The rest of the difference we attribute to differencies between the
numerical codes (hydro vs. N-body).

\begin{figure}[h]
\centerline{\includegraphics[scale=0.58]{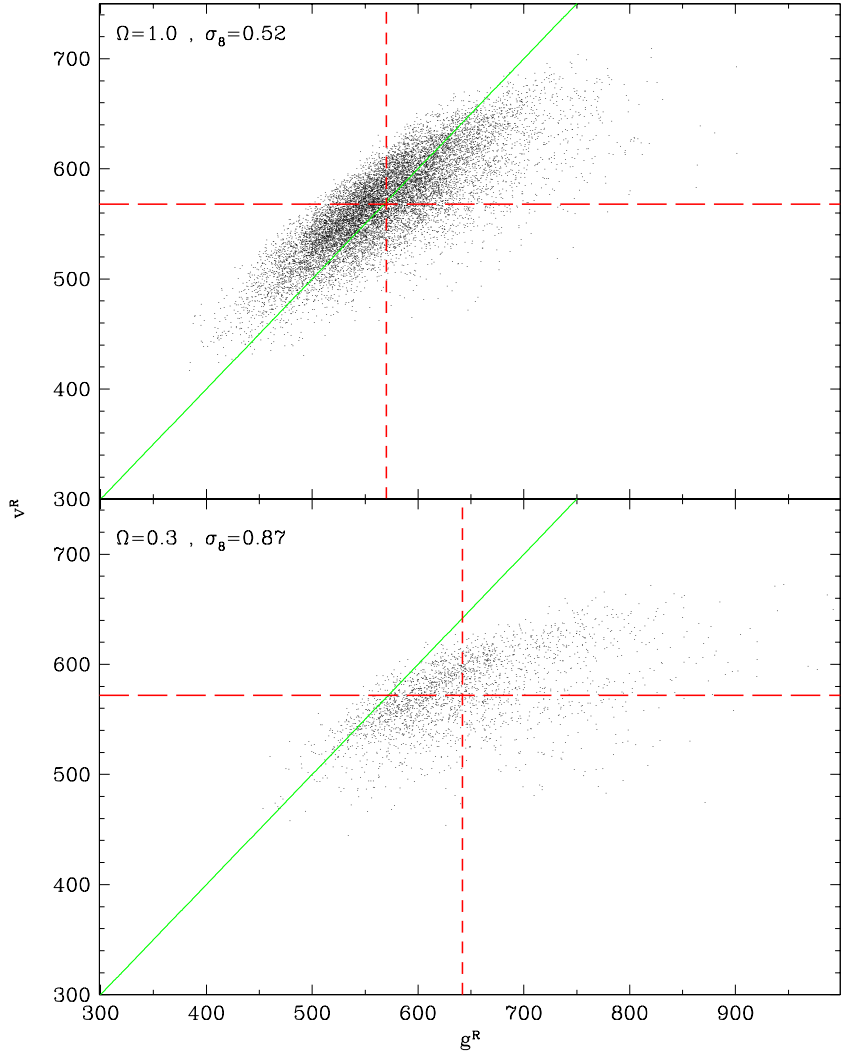}}
\caption{Modules of 500~\kms ~top-hat smoothed velocity and gravity 
for the LG candidates (scatter plots). 
Top panel: $\Omega =1$, bottom: $\Omega =0.3$. 
Solid line represents the linear relation, 
dashed -- mean values of $g^R$ and $v^R$.
\label{fig:VG_LG}
}
\end{figure}

\Section{The relation for amplitudes}
\label{sec:am}
In the previous Section we have shown that the misalignment angle
between the vectors of velocity and gravitational acceleration is
small. Therefore, most information about the relation between the two
vectors is contained in the relation between their amplitudes.
This is also the crucial point in determining $\Omega$ from the
LG dipole. Figure~\ref{fig:VG_LG} shows this relation for the LG 
candidates in models with $\Omega = 1$ and $\Omega = 0.3$. 
Here, both gravity and velocity fields are smoothed with a top-hat 
filter of radius $500$ \kms.

In the top panel of Figure~\ref{fig:VG_LG}, the points agree with the 
linear prediction (solid line) quite well. 
The reason of this is twofold. First, the
model with $\Omega =1$ has a low normalization, so the velocity and
gravity fields are only weakly nonlinear. Second, the velocity $600$
\kms\ is rather typical for an $\Omega = 1$ universe, while the
velocity--gravity relation deviates more significantly from the linear
prediction only in the high-velocity tail (Kudlicki \etal
2000b). Specifically, the r.m.s.\ value of the velocity of all points
in the simulation is $458$ \kms. 

Contrary to the previous case, in the lower panel of
Figure~\ref{fig:VG_LG} the points lie farther off the line.  This is
not surprising, because the normalization of the model with $\Omega =
0.3$ is higher, so the fields are more nonlinear. Moreover, the LG
velocity of $600$ \kms\ is less typical for an $\Omega = 0.3$
Universe. Higher normalization of the model compensates for this
effect only partly. The r.m.s.\ value of the velocity of all points in
this model is $340$ \kms.

\begin{figure}[h]
\centerline{\includegraphics[angle=270,scale=0.7]{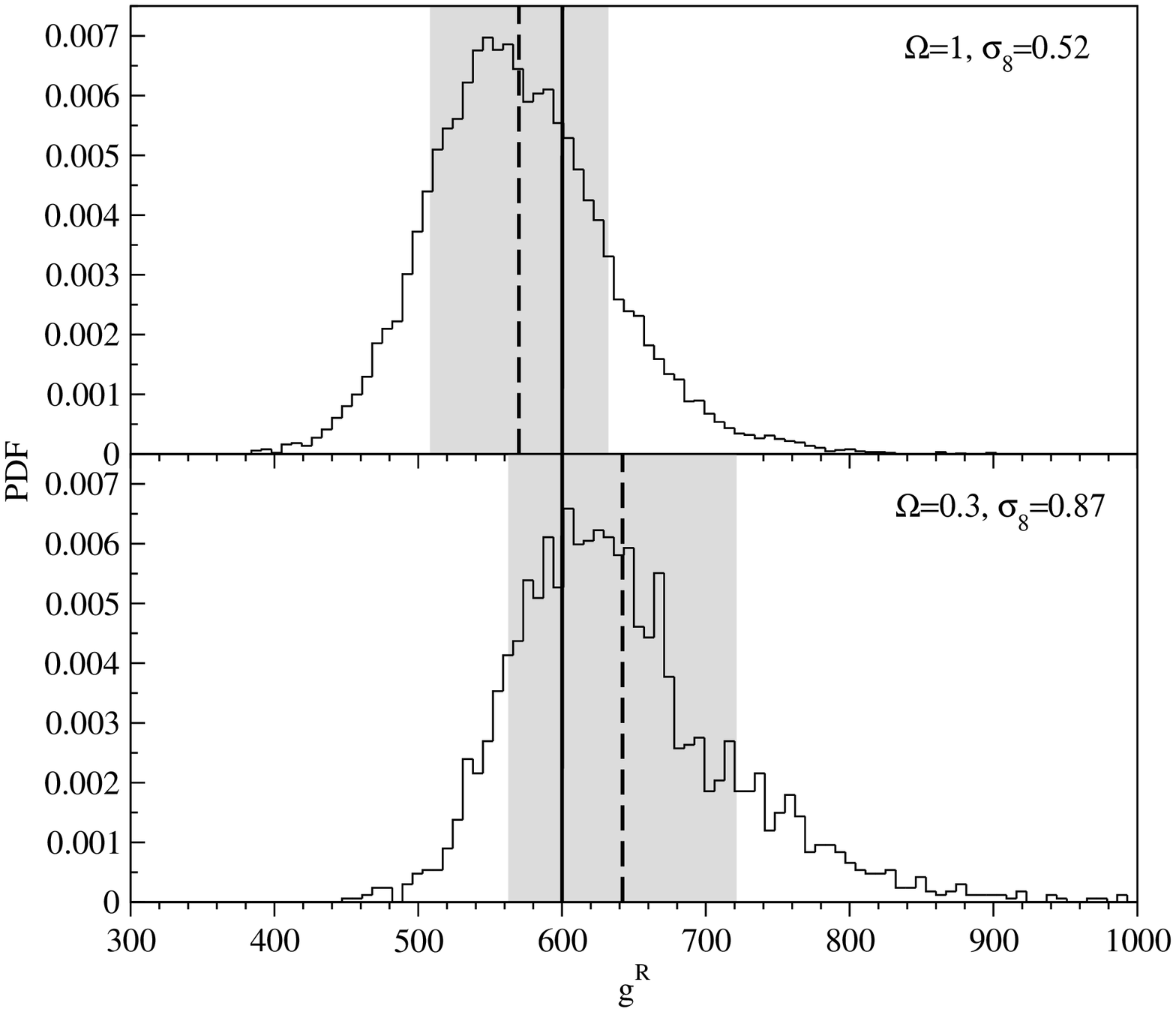}}
\caption{ PDF of the smoothed gravity field (histogram). 
Solid vertical line marks the value of 600 \kms, dashed -- mean values 
of $g^R$ and shaded strips -- 1$\sigma$ intervals around the mean values.
\label{fig:GHIST_LG}
}
\end{figure}

The above discussion does not, however, imply that the gravitational
acceleration of the LG is a better estimator of the LG velocity in a
$\Omega = 1.0$ than in a $\Omega = 0.3$ Universe. As described
earlier, in actual comparisons one uses the smoothed gravity of the LG
as an estimator of its {\em unsmoothed\/}\footnote{Strictly speaking,
smoothed over the small volume of the LG.} velocity. In
Figure~\ref{fig:VG_LG} we see that, for both models, the mean
smoothed velocity of the LG candidates (dashed horizontal line) is
smaller than their unsmoothed velocity, $v_{} = 600$ \kms. As stated
in the previous Section, it means that part of the central velocity
has been induced by mass fluctuations within the smoothing
radius. Since $\lan v^R \ran < v_{}$, though for the flat model $\lan
v^R \ran \simeq \lan g^R \ran$, $\lan g^R \ran$ should be smaller than
$v_{}$. Figure~\ref{fig:GHIST_LG} show distributions of the values of $g^R$. 
For $\Omega=1$, $\lan g^R \ran = 570$ \kms(dashed vertical line),
so indeed $\lan g^R \ran < v_{}$. Thus, the smoothed gravitational
acceleration of the LG is in this case a slightly biased estimator of
its unsmoothed velocity, and essentially all bias results from
estimating unsmoothed velocity by its smoothed counterpart, and not
from nonlinear gravity. The difference between $\lan g^R \ran$ and
$v_{}$ results in a systematic error of the estimate of $\beta$ of
5.3 \%.

As for the model with $\Omega = 0.3$, the situation is different. The
effects of nonlinear gravity act in the opposite direction than the
effects of smoothing and turn out to dominate over them. The mean
value of $g^R$ is here $642$ \kms. In other words, in the $\Omega =
0.3$ model the smoothed gravitational acceleration of the LG {\em
overestimates\/} its unsmoothed velocity. This results in a systematic
error of the estimate of $\beta$ of $6.5$ \%.

To sum up, smoothed gravity of the LG underestimates its velocity in a
high-$\Omega$ Universe, while it overestimates the velocity in a
low-$\Omega$ Universe. Thus, there is no common bias, independent of
the assumed model, that can be corrected for in the data analysis.
Therefore, the results should be either expressed in a model-dependent
way or they are a subject to this systematic error. Fortunately, this
error is in any case relatively small, of a few per cent.

Thus far, we have left aside the scatter in the velocity--gravity
relation. This scatter results in a random error in the estimate of
$\beta$ from the LG velocity--gravity comparison. Unlike the
systematic error, this error cannot be reduced in any model, because
in practice one does not have at his disposal a whole set of the LG
observers, but merely one observer (us). Thus, the NE put also an
upper limit on the {\em precision\/} of this method of determining
$\beta$. The dispersion of the values of $g^R$, $\de g^R$, is $62$ and
$79$ \kms, respectively for the models with $\Omega =1$ and $\Omega =
0.3$. This implies that the $1\sigma$ random error in the estimate of
$\beta$ due to the NE is $12.3$ \% and $10.9$ \%, respectively.

The summary of systematic and random errors is presented 
in Table~\ref{tab:beta}.

\begin{table}[h]
\centerline{
\begin{tabular}{c|c|c}
$\Omega$ & $v/\lan g^R \ran - 1$ & $ \de g^R/\lan g^R \ran $ \\
\hline
0.3      &   -6.5 \%    &    10.9 \% \\
\hline
1.0      &  5.3 \%    &    12.3 \% \\
\end{tabular}
}
\caption{Systematic and random errors in determining
$\beta$ from the LG dipole comparisons due to
non-linear effects, for two cosmological models.
\label{tab:beta}
}
\end{table}

\Section{Summary}
\label{sec:sum}
Using numerical simulations we have studied the relation between the
velocity of the Local Group and its gravitational acceleration. This
relation underlies both a test for the kinematic origin of the CMB
dipole and a method of estimating $\beta$.

First, we have investigated the mean misalignment angle between the
two vectors. For the case of $\Omega =1$, we have confirmed the result
of Davis \etal (1991), that this angle is small. Actually, the
value we have obtained ($\sim 7^\circ$) is even smaller than that of
Davis \etal ($\sim 10^\circ$). We attribute a part of the difference
to the different power spectrum we used, namely, that of the PSCz
survey.  The rest of the difference is likely caused by the
differences between the numerical codes (hydro vs.\ N-body). We have
verified that the misalignment angle is fairly insensitive to
$\Omega$, in agreement with our theoretical prediction.

The {\em observed} misalignment angle is $\sim 14^\circ$, 
close to the upper limit for the $95$ \%
confidence interval for the simulated angle. Therefore, the NE are
unlikely to be responsible for all of the observed misalignment. Other
effects must play a role here as well, like, e.g., incomplete sky
coverage and/or shot noise.

Next, we have studied the relation between the amplitudes of the LG
velocity and gravity vectors. This relation has a scatter,
resulting in a $11$--$12$ \% random error of the estimate of $\beta$. 
In the model with $\Omega = 1$, the smoothed gravity of
the LG turns out to be a biased low estimator of the LG velocity. On
the contrary, in the model with $\Omega = 0.3$, the estimator is
biased high.

Using mock catalogs constructed for different cosmologies, S99
computed the ratio between the reconstructed gravity dipole at the
observer's position and its true N-body velocity. In a `numerical
analysis', they subsequently used the average of all ratios as a
multiplicative factor which relates the reconstructed gravity of the
LG to the true LG velocity. Since biases due to the NE are distinct
for different cosmologies (and this is also true for other effects,
like, e.g., the effect of finite volume), an attempt to derive a
model-independent estimate of $\beta$ is illegitimate. Rather, the
results should be expressed explicitly in a model-dependent
way. Indeed, the estimates of $\beta$ for models with $\Omega = 1$ and
with $\Omega = 0.3$, obtained by S99 using an alternative,
model-dependent maximum-likelihood analysis are clearly different. We
conclude that the `numerical analysis' of S99 is formally incorrect.

While in principle important, in practice the discussed biases are
small: of the order of a few per cent. With this accuracy, therefore,
to the LG velocity--gravity comparisons the linear theory can be 
applied. Furthermore, the random error of the estimate of $\beta$ due
to the nonlinear effects is small compared to the total random error
($\sim$ 40\%: S99). We thus conclude that in the LG velocity--gravity
comparisons the NE are not the major concern.

\Acknow{
We thank Micha{\l} R\'o\.zyczka for useful comments. This
research has been supported in part by the Polish State Committee for
Scientific Research grants No.~2.P03D.014.19 and 2.P03D.017.19. The
numerical computations reported here were performed at the
Interdisciplinary Centre for Mathematical and Computational Modeling
in Warsaw, Poland.}

\end{document}